\input epsf
\documentstyle[preprint,prd,aps,rotate]{revtex}
  
\begin{document}

\title{SPHERES -- OMNI-DIRECTIONAL MULTI-MODE GRAVITATIONAL-WAVE ANTENNAS
FOR THE NEXT GENERATION }
\author{Ho Jung Paik\footnote{%
e-mail: h\_paik@umail.umd.edu}}
\address{Department of Physics, University of Maryland, \\College Park,
Maryland 20742, U.S.A. }
\maketitle

\begin{abstract} 
The sensitivity of resonant-mass gravitational-wave antennas has now reached 
$h \leq  10^{-18}$, where $h$ is the dimensionless metric perturbation
caused by the wave. In  order to observe gravitational wave events such as
coalescing neutron star binaries and  colliding black holes from as far as
the Virgo cluster of galaxies, the detector sensitivity  must be improved by
three orders of magnitude. With such an aim, construction of  massive ($\sim
40$ ton) spherical antennas have been proposed by a number of groups world 
wide. The standard quantum limit of the sensitivity of such detectors will
be $h \cong 3  \times 10^{-22}$. Unlike a cylinder, a sphere has five
degenerate quadrupole modes  which interact with an incoming gravitational
wave. By combining the responses of all  these modes, one can determine the
direction of the source and the polarization of the  wave with a single
spherical antenna. Summed over the modes, the sphere has a 
direction-independent absorption cross section, as expected from its
symmetry, permitting full-sky  coverage with a single antenna. These unique
properties, combined with its enhanced  sensitivity due to its multi-mode
nature and increased mass, make spherical detectors ideal new instruments
for observational astronomy. Significant  advances have been made recently
on instrumentation of a spherical antenna. Mounting  six identical resonant
transducers on the six pentagonal faces in one hemisphere of a  truncated
icosahedral gravitational-wave antenna (TIGA) has been found to maintain the
``spherical" symmetry in the coupled antenna-transducer system. The
quadrupole modes of  the sphere split into doublets with equal frequency
splitting, as desired. Numerical  calculations have shown that
gravitational-wave signals from coalescing neutron star  binaries
in the Virgo cluster can be 
resolved with a spherical antenna with a near-quantum-limited sensitivity. 
\end{abstract}

\vskip 1cm \noindent {\bf I. Introduction} \vskip 0.7cm

It has been known for eight decades that Einstein's theory of general
relativity predicts the existence of gravitational waves \cite{Einstein}.
According to general relativity, the gravitational wave (GW) is transverse
with spin 2 and travels at the speed of light. These waves are expected to
be emitted by cosmic events such as supernova explosions, collisions of
black holes, and coalescence of compact binaries. Interacting weakly with
matter, GWs are expected to penetrate regions of dense star population,
which block electromagnetic waves of all frequencies. Therefore, GW
detectors will be a powerful new tool for studying many interesting
astrophysical phenomena which may otherwise remain unobservable.

The extreme weakness of GW signals from extragalactic sources, however,
poses a major challenge for the construction of instruments sensitive enough
to detect the predicted astronomical events. Two types of GW detectors are
under development: resonant-mass detectors and laser interferometers \cite
{review}. The resonant-mass detectors have gone through the first (300 K,
cylindrical) and second generations (4.2 K, cylindrical), and now
third-generation detectors (50 mK, cylindrical) are being tested \cite
{Nautilus,Auriga}. Development of laser interferometers is going faster.
After prototyping at a $30\sim 40$ m baseline, long-baseline ($3\sim 4$ km)
interferometers are under construction \cite{LIGO,VIRGO}. There is now a
new initiative in the resonant-mass detector community to build large
spherical antennas which will be cooled to 50~mK or below, fourth-generation
resonant-mass antennas.

Spherical antennas have the advantage of having a uniform absorption cross
section independent of  source direction and polarization
(omni-directionality) and of being able to detect the  source direction and
polarization with a single antenna \cite{WagonerPaik}. In their  somewhat
restricted bandwidths, the sensitivity of the large ($40 \sim 100$ tons)
spherical  detectors operating near the standard quantum limit will be
comparable to that of  long-baseline laser interferometers. By operating the
two types of detectors simultaneously, the  accidental coincidence rate can
be reduced, the signal-to-noise ratio for GW signals  can be improved, the
signal parameters can be better determined. Therefore, the spheres  will be
a good complement to the laser interferometers under construction.

In this paper, we will first review the basic properties of cylindrical
antennas and  summarize the theory of resonant-mass detectors. We will then
present the concept of  spherical detectors and discuss some astrophysical
events that could be detected with these  new detectors.

\vskip 1cm \noindent {\bf II. Basic Properties of a Cylindrical Antenna} %
\vskip 0.7cm

Most resonant-mass antennas constructed thus far have been cylinders of
aluminum alloy,  so-called ``Weber bars.'' Their typical dimensions are $L
\cong 3$ m (length) and $R  \cong 0.3$ m (radius), which give $f_{0} \cong
900$ Hz (fundamental longitudinal  frequency) and $M_{tot} \cong 2.3$ tons
(total mass).

Let us define the coordinates of the antenna in the wave frame such that the
GW is purely  ``+" polarized and the wave travels in the $z$ direction. The
origin of the coordinate system  is located at the geometric center of the
cylinder. Let $\theta$ and $\phi$ represent the  polar angles of the
cylinder axis. Then the energy absorbed by the $n$-th longitudinal  harmonic
of the antenna for $n$ odd can be written \cite{MTW} as 
\begin{equation}
E_{n} =  \frac{M_{tot}v^{4}}{L^{2}} {|h(\omega_{n})|}^{2}
\sin^{4} \theta \cos^{2} 2 \phi  \label{eq:EnergyAbsorbed}
\end{equation}
where $v$ is the speed of sound in the antenna material and $h(\omega)$ is
the Fourier  component of the dimensionless metric perturbation $h(t)$ at
(angular) frequency $\omega$.  For $n$ even, $E_{n}$ vanishes. The angular
factor shows the directionality of the  antenna. The cylindrical antenna has
a maximum sensitivity for signals arriving  perpendicular to its axis but
misses signals traveling along its axis.

It is convenient to write the absorbed energy in terms of a dimensionless 
{\it reduced cross  section} $\Pi$: 
\begin{equation}
E = \Pi \frac{\pi \rho v^{5}}{2f_{0}} {|h(\omega)|}^{2}
\label{eq:EnergyReduced}
\end{equation}
From Equations \ref{eq:EnergyAbsorbed} and \ref{eq:EnergyReduced}, one can 
identify the reduced cross section for the fundamental ($n = 1$)
longitudinal mode of a cylinder as 
\begin{equation}
\Pi_{C} = (R/L)^{2} \sin^{4} \theta \cos^{2} 2 \phi
\label{eq:ReducedCylinder}
\end{equation}
Therefore, $\Pi_{C} \cong 0.01$ for optimal source orientation and
polarization. Upon  all-sky average, the angular factor becomes 4/15 so that
the {\it average} reduced cross  section of a cylindrical antenna is $%
<\Pi_{C}> \cong 0.002$.

\vskip 1cm \noindent {\bf III. Theory of Resonant-Mass Gravitational-Wave
Detectors} \vskip 0.7cm

\begin{figure}[tbp]
\begin{center}
\leavevmode
\epsfxsize=3.5in\epsfbox{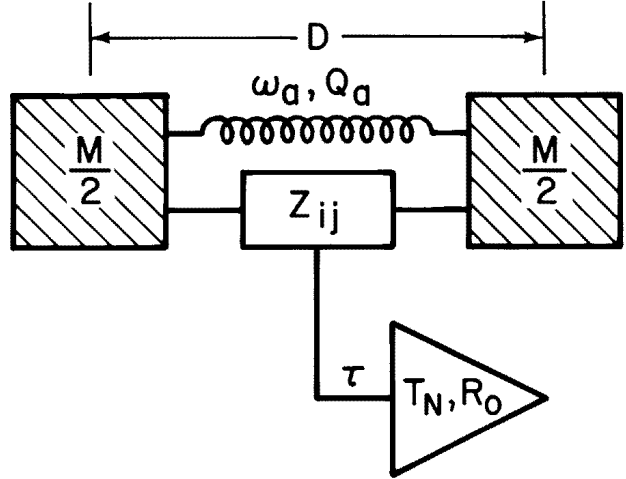}
\end{center}
\caption{Schematic of a resonant-mass GW detector}
\label{fig:DetectorModel}
\end{figure}

A GW detector is schematically shown in Figure \ref{fig:DetectorModel}. The
antenna is  parameterized in terms of its effective mass $M$, length $L$,
(angular) resonance  frequency $\omega_{a}$, and quality factor $Q_{a}$. The
transducer is characterized by  its impedance matrix $Z_{ij}$, which relates
the input velocity $u(\omega)$ and the  output current $I(\omega)$ with the
input force $f(\omega)$ and the output voltage $V(\omega)$ by 
\begin{equation}
\left( 
\begin{array}{c}
f(\omega) \\ 
V(\omega)
\end{array}
\right) = \left( 
\begin{array}{cc}
Z_{11} & Z_{12} \\ 
Z_{21} & Z_{22}
\end{array}
\right) \left( 
\begin{array}{c}
u(\omega) \\ 
I(\omega)
\end{array}
\right)  \label{eq:ImpedanceMatrix}
\end{equation}
The amplifier is characterized by its noise temperature $T_{N}$, optimum
source  impedance $R_{0}$, and integration time $\tau$.

\vskip 3cm {\bf 1. Detector Response to Signal and Noise} \vskip 0.7cm

Let us consider a short pulse of GW which consists of one cycle of a sine
wave with  amplitude $h$ and (angular) frequency $\omega_{S} \, (\cong
\omega_{a})$. The energy  deposited by this pulse into a favorably oriented
cylindrical antenna at rest can be shown  to be 
\begin{equation}
E_{S} \approx \frac{2}{\pi^{2}} M {\omega_{S}}^{2} {(Lh)}^{2}
\label{eq:SignalEnergy}
\end{equation}
A careful treatment of all the noise terms of the detector \cite{Giffard}
leads to the total  noise energy 
\begin{equation}
E_{N} \approx \frac{1}{2} k_{B} T_{a} \frac{\omega_{a} \tau}{Q_{a}} + \frac{1%
}{2} k_{B} T_{N} \left[ \frac{2(\zeta + \zeta^{-1})}{\beta_{21} \omega_{S}
\tau} + \frac{\beta_{12} \omega_{S} \zeta \tau}{2} \right]
\label{eq:NoiseEnergy}
\end{equation}
where $\beta_{21} \equiv {|Z_{21}|}^{2} / (M \omega_{S} |Z_{22}|)$ is the 
{\it \ forward} energy coupling coefficient, $\beta_{12} \equiv {|Z_{12}|}%
^{2} / (M  \omega_{S} |Z_{22}|)$ is the {\it reverse} energy coupling
coefficient, $\zeta \equiv  |Z_{22}|/R_{0}$ is the dimensionless impedance
matching parameter, and $\tau \approx {(2 \Delta f_{S})}^{-1}$ is the
integration time. Here $\Delta f_{S}$ is the signal  bandwidth of the
detector. We will confine our discussions to {\it passive} transducers,  for
which $\beta_{12} = \beta_{21} \equiv \beta_{S}$.

The two terms in Equation \ref{eq:NoiseEnergy} represent the Brownian noise
of the  antenna-transducer system and the amplifier noise, respectively. Of
the amplifier noise,  the first term in the square brackets is due to the
wide-band noise appearing at the output  of the amplifier and the second
term represents the so-called ``back-action noise," i.e., the  amplifier
noise fed back to the antenna. The Brownian motion noise can be reduced to
an  arbitrarily low level, in principle, by reducing $T_{a}/Q_{a}$. The
amplifier noise,  however, does have a fundamental limit. The noise
temperature of {\it linear} amplifiers  has a quantum limit, $T_{N} \geq
\hbar \omega_{S} / k_{B}$ \cite{Heffner}.

\vskip 1cm {\bf 2. Optimization of Detector Parameters} \vskip 0.7cm

Optimization conditions for detector parameters can be obtained as follows.
Since the  two terms in the square brackets of Equation \ref{eq:NoiseEnergy}
are proportional to $\tau^{-1}$ and $\tau$, respectively, $\tau$ must be
chosen such that the two terms become  equal, in order to minimize the total
amplifier noise contribution (although the shortest $\tau$ is desirable to
minimize the Brownian motion noise term). This leads to 
\begin{equation}
\tau = \frac{2}{\beta_{S} \omega_{S}} \sqrt{1 + \frac{1}{\zeta^{2}}}
\label{eq:OptimumIntegration}
\end{equation}
The impedance matching parameter must satisfy $\zeta \leq 1$ in order to
avoid blowing  up any of the amplifier noise terms. With this choice,
Equation \ref{eq:OptimumIntegration} implies $\Delta f_{S} / f_{S} \approx
\beta_{S}$. This is a  very important result. The fractional bandwidth of a
resonant-mass antenna can approach  unity {\it if} a near-unity $\beta_{S}$
can be achieved. A {\it multi-mode} transducer  scheme can satisfy this
condition, and will be discussed in the next section. The condition that the
antenna-transducer system must satisfy in order for its Brownian motion
noise to become  negligible compared to the amplifier noise is given by 
\begin{equation}
T_{a}/Q_{a} \ll \beta_{S} T_{N}  \label{eq:TemperatureCondition}
\end{equation}

A wide-band detector thus reduces the $T_{a}/Q_{a}$ requirement. If this
condition is  satisfied, the detector noise reaches the amplifier noise
limit: $E_{N} \approx k_{B}  T_{N}$. Finally, combining this with Equation 
\ref{eq:SignalEnergy}, one obtains the  minimum detectable signal: 
\begin{equation}
h_{min} \approx \left( \frac{5k_{B} T_{N}}{M {\omega_{S}}^{2} L^{2}}
\right)^{1/2} \geq \left( \frac{5 \hbar}{M \omega_{S} L^{2}} \right)^{1/2}
\label{eq:QuantumLimit}
\end{equation}
The last expression here represents the {\it standard quantum limit} of the
detection sensitivity.  For the typical cylindrical antenna whose dimensions
were given in Section II, the standard  quantum limit corresponds to $%
h_{min} \approx 3 \times 10^{-21}$. So far second-generation  resonant-mass
detectors have reached about 100 times this level, due to their relatively
high  operating temperature, $T_{a} = 4.2$ K, and limited amplifier noise, $%
k_{B} T_{N}/ \omega_{S}  \geq 10^{4} \hbar$.

\vskip 1cm {\bf 3. Multi-mode Transducer} \vskip 0.7cm

As the amplifier noise approaches the quantum limit, $T_{N, QL} = 4 \times
10^{-8}$ K,  the $T_{a}/Q_{a}$ requirement for the antenna-transducer system
becomes very severe.  At the temperature of the third- and fourth-generation
antennas, $T_{a} = 0.05$ K, $Q_{a}  \gg 10^{7}$ is required, even with $%
\beta_{S} = 0.1$, in order to reach the sensitivity  given by Equation \ref
{eq:QuantumLimit}. Therefore, a wide-band transducer is of  crucial
importance.

\begin{figure}[tbp]
\begin{center}
\leavevmode
\epsfxsize=3.5in\epsfbox{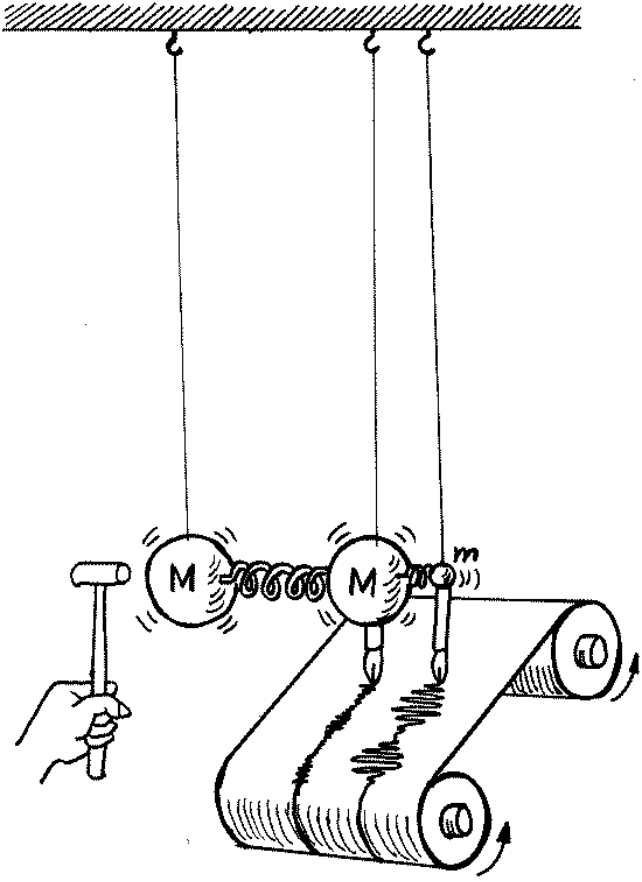}
\end{center}
\caption{Principle of a resonant transducer}
\label{fig:ResonantTransducer}
\end{figure}

A {\it resonant} transducer has been introduced as the first step toward
wide-band  detection \cite{Paik}. Its principle is illustrated in Figure \ref
{fig:ResonantTransducer}.  A small resonant mass $m \, (\ll M)$ is attached
to the antenna with effective mass $M \,  (= M_{tot}/2$ for cylinders). Then
the frequency splits into $f_{+}$ and $f_{-}$, and the  energy flows back
and forth between the antenna and the transducer with a beat frequency: 
\begin{equation}
f_{B} = f_{+} - f_{-} = f_{a} \sqrt{m/M}  \label{eq:Beat}
\end{equation}
The amplitude gain of the transducer, $\sqrt{M/m}$, leads to improvement of $%
\beta_{S}$ roughly by a ratio of $M/m$. There is an optimum transducer mass, 
$m_{opt}$, since too small a transducer mass results in an unacceptably long
beat period  which compromises $\tau$. Since $\tau \approx 1/(2f_{B})$, one
finds 
\begin{equation}
\beta_{S} \approx \Delta f_{S}/f_{S} \approx \sqrt{m_{opt}/M}
\label{eq:TwomodeBandwidth}
\end{equation}

The fractional bandwidth of the detector can be widened further by adding
intermediate  resonant masses between the antenna and the final transducer
mass \cite{Richard}. For an $N$-mode system with a constant mass ratio
between the neighboring masses,  Equation \ref{eq:TwomodeBandwidth} is
generalized into 
\begin{equation}
\beta_{S} \approx \Delta f_{S}/f_{S} \approx (m_{opt}/M)^{1/2(N-1)}
\label{eq:MultimodeBandwidth}
\end{equation}
In principle, the fractional bandwidth can be increased arbitrarily close to
unity by  increasing $N$. In practice, increasing $N$ complicates the
construction and operation of  the detector, and the bandwidth gain is slow
beyond $N = 3$. Thus a three- or four-mode  system will be a practical limit
in most cases. A three-mode ($N = 3$) detector with $m_{opt}/M = 10^{-4}$
gives $\beta_{S} \approx 0.1$.

\vskip 1cm \noindent {\bf IV. Spherical Antennas of Gravitational Waves} %
\vskip 0.7cm

The sensitivity reached with 4 K cylindrical detectors appears to be good
enough to detect  supernova events and coalescing neutron star binaries in
our galaxy. However, the  expected rates for these events are only one every
50 years or less \cite{Thorne}. In order  to improve the event rates to
about one a year, one needs to look all the way to the Virgo  cluster of
galaxies. This requires improvement of the detection sensitivity by another
three  orders of magnitude in amplitude, to $h_{min} \approx 3 \times
10^{-22}$, which is a  factor of 10 beyond the standard quantum limit for
the cylinders! Although back-action  evasion is possible in principle to
beat the standard quantum limit \cite{Caves}, the  engineering challenge of
reducing the thermal noise further in the antenna-transducer  system makes
it very difficult to realize in practice. Achieving a larger antenna cross
section by using a  larger effective mass or a material with a higher speed
of sound may be a better approach.  As will be shown, a {\it spherical}
antenna should be able to provide the last factor of 10 improvement  in
sensitivity.

\vskip 1cm {\bf 1. Advantages of a Spherical Antenna} \vskip 0.7cm

A spherical antenna was originally proposed by Forward \cite{Forward} as a
means of  detecting both the scalar and tensor gravitational waves predicted
by various modern theories of  gravity. The scalar wave would excite only
the monopole ($\ell = 0$) mode of the sphere  whereas the tensor wave would
interact with only the quadrupole ($\ell = 2$) modes. Since there  are five
degenerate quadrupole modes to observe $(\psi_{2,2}$, $\psi_{2,1}$, $%
\psi_{2,0}$, $\psi_{2,-1}$, $\psi_{2,-2})$ while there are only four
unknowns for a GW signal, the two  source angles $(\theta, \phi)$ and the
amplitudes for the two polarizations $(h_{+},  h_{\times})$, all these
parameters could be determined with a single spherical antenna and  the
remaining fifth degree of freedom could be used to veto non-GW events.

Wagoner and Paik \cite{WagonerPaik} computed the absorption cross section of
a sphere  and solved the inverse problem of determining the direction and
polarization of the wave  from the mode excitations. In particular, they
found that, for the same total mass, a  sphere has approximately five times
the cross section of a cylinder averaged over the  direction and the
polarization. On the other hand, for the same antenna frequency $f_{a}$  and
the same material, a sphere has roughly 20 times more mass than a typical
cylinder.  Thus the reduced cross section of a sphere is about 100 times
greater than that of a  cylinder: 
\begin{equation}
\Pi_{S} = 0.215  \label{eq:ReducedSphere}
\end{equation}
This improves the amplitude sensitivity by a factor of 10, as required.
Further, unlike  the cylinder, the reduced  cross section of a sphere is
independent of the source direction. Thus a single antenna  provides
full-sky coverage with equal sensitivity and determines the source direction
and  the polarization \cite{ZhouMichelson}. A coincidence experiment across
two identical  spheres will give coincidence not only in time, but also in
the direction and the  polarization, and in the amplitude of excitation, a
very strong criterion for coincidence!  The sphere is indeed an {\it ideal}
GW detector.

Another interesting property of a sphere is that higher
harmonics of a larger sphere have significantly  bigger cross sections than
the fundamental modes of smaller spheres resonant at the same  frequencies 
\cite{Lobo}. Therefore, the spheres could be instrumented as multi-frequency
antennas.

\vskip 1cm {\bf 2. TIGA Arrangement for Transducers} \vskip 0.7cm

\begin{figure}[tbp]
\begin{center}
\leavevmode
\epsfxsize=3.5in\epsfbox{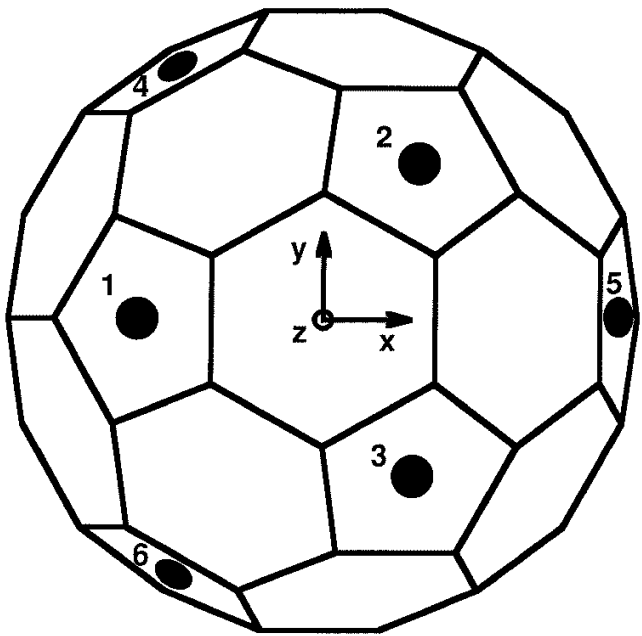}
\end{center}
\caption{Truncated icosahedral gravitational-wave antenna}
\label{fig:TIGA}
\end{figure}

But how could one instrument the sphere with resonant transducers without
destroying the  degeneracy of the modes and disturbing the
omni-directionality of the sphere? This  important question was answered in
recent papers by Johnson and Merkowitz \cite{Johnson,Merkowitz}. They
proposed the so-called ``TIGA" (Truncated Icosahedral  Gravitational-wave
Antenna) arrangement in which six radial transducers are mounted on  six
pentagonal surfaces in a hemisphere of a truncated icosahedron. These
locations are  shown by black dots in Figure \ref{fig:TIGA}. (The pentagonal
surfaces alone form a  dodecahedron.)  The proposed  six locations give the
maximum isotropy and the resulting dodecahedron has the same symmetry as
the sphere up to the quadrupole moment.

If the transducers are tuned to the frequency of the fundamental quadrupole
modes  of the sphere, the five antenna and the six transducer modes interact
with each other to  produce five degenerate modes at an upper frequency $%
f_{+}$, five degenerate modes at a  lower frequency $f_{-}$, and one mode at
the unshifted frequency $f_{a}$. The beat  frequency between the $f_{+}$ and 
$f_{-}$ modes is given by Equation \ref{eq:Beat}  with $M = 0.26 M_{tot}$
for a sphere. The last mode corresponds to the monopole combination of  the
transducer outputs which does not interact with the sphere quadupole modes. 
Therefore, there must be five independent linear combinations of the
transducer outputs of  which each interacts uniquely with each of the five
sphere modes.

Merkowitz and Johnson derived these linear combinations, called ``mode
channels" \cite{Merkowitz}. To do this, they first defined a new set of
orthogonal sphere modes,  described by {\it real} functions $\psi_{m}$, such
that 
\begin{equation}
\begin{array}{l}
\psi_{1} \equiv \frac{1}{\sqrt{2}}(\psi_{2,-2} + \psi_{2,2}) \\ 
\psi_{2} \equiv \frac{1}{\sqrt{2}}(\psi_{2,-2} - \psi_{2,2}) \\ 
\psi_{3} \equiv \frac{1}{\sqrt{2}}(\psi_{2,-1} + \psi_{2,1}) \\ 
\psi_{4} \equiv \frac{1}{\sqrt{2}}(\psi_{2,-1} - \psi_{2,1}) \\ 
\psi_{5} \equiv \psi_{2,0}
\end{array}
\label{eq:RealModes}
\end{equation}
Let $q_{j}(t)$ be the displacement response of the $j$-th transducer. The
mode channels $g_{m}(t)$  are then defined as 
\begin{equation}
g_{m}(t) \equiv \psi_{m}(R, \theta_{j}, \phi_{j}) q_{j}(t)
\label{eq:ModeChannel}
\end{equation}
Each transducer mode channel couples only to the corresponding sphere mode,
and the  antenna-transducer system behaves just like five independent
two-mode cylindrical  antennas. The optimal filtering algorithm developed
for cylindrical antennas can be  therefore be applied to each mode channel.
The direction and polarization of the wave  can be solved from the
determined amplitudes of the sphere modes by using the method  prescribed by
Wagoner and Paik \cite{WagonerPaik}.

Another question that must be addressed is how one should suspend the sphere
to  maintain its symmetry along with its other nice properties. A nodal
support from near its  center of mass is certainly an attractive solution,
which also provides additional vibration  isolation. A multi-point support
from the surface of the sphere should also work if the  support structure
has its lowest eigenfrequencies low compared to the antenna frequencies  and
the support cables are carefully tuned to reflect the sound waves at the
antenna  frequencies coming down from the room-temperature end.

\vskip 1cm \noindent {\bf V. Development Program for Spherical Detectors} %
\vskip 0.7cm

With the advantages of spherical detectors clearly understood and with the
questions on  the transducer arrangement satisfactorily answered, the
resonant-mass detector groups throughout the  world are joining their
efforts to build large ultra-cryogenic spherical detectors. A U.S.
consortium consisting  of research groups at Louisiana State University,
University of Maryland, University of  Rochester, and Santa Clara University
has submitted to the National Science Foundation a  joint R\&D proposal for
TIGA. If successful, this will lead to construction of large  spherical
antennas in the following few years. A Dutch consortium led by researchers
at the  University of Leiden has started R\&D for a spherical antenna called
GRAIL (Gravitational  Radiation Antenna In Leiden). Groups in Italy and
Brazil are also conducting research on  spherical detectors. All these
groups have joined together to form an international  collaboration under
the project name of OMEGA (OMni-directional Experiment with 
Gravitational-wave Antennas).

The present concept of the TIGA project is to build a ``xylophone" of four
aluminum alloy  spheres with diameters ranging from 2 to 3 meters. The
largest sphere will then weigh  about 40 tons and have the lowest quadrupole
frequency of about 900 Hz. A three-mode  antenna-transducer system will
allow a fractional bandwidth of 0.1. If both the  fundamental and second
harmonic quadrupole modes are instrumented, a frequency range  of 800 to
2700 Hz will be be covered. The antennas will be cooled to 50 mK by  He$^{3}$%
/He$^{4}$ dilution refrigerators. The baseline transducer for TIGA is a 
superconducting inductive transducer \cite{Paik} coupled to a
near-quantum-limited dc  SQUID (Superconducting QUantum Interference
Device). As a backup, an optical  transducer based on a Fabry-Perot
interferometer is also under development \cite{RichardPang}.

A denser, higher sound speed material is desirable for the antenna in order
to increase its  cross section. However, aluminum 5056 has been chosen
tentatively because  of its manufacturability into large spheres and its
proven high mechanical quality factor.  Explosive bonding techniques will
permit construction of a 40 ton aluminum sphere.  Samples of explosively
bonded aluminum 5056 has been tested with good  success. After annealing, a
Q in excess of $3 \times 10^{7}$ was measured in a torsional  oscillator
made out of the explosively bonded material \cite{Duffy}. This is very close
to  the highest Q ever measured in the monolithic material of aluminum 5056,
and is within a  factor of 2 from the value required for the standard
quantum limit.

The superconducting transducer is electrically simple due to its passive
nature and has  operated reliably on a number of cryogenic antennas. It
consists of just superconducting  coils carrying a persistent current and a
dc SQUID. Displacement of a superconducting  test mass with respect to the
coils modulates the persistent current and produces a  time-varying magnetic
flux which is detected by the SQUID. Significant progress has been  made
recently in the fabrication of a practical dc SQUID with low noise. A
well-coupled dc  SQUID with flux noise of about $50 \hbar$ at a temperature
below 100 mK has been  demonstrated by Wellstood's group at the University
of Maryland \cite{Wellstood}. A new SQUID with expected noise  below $10
\hbar$ has been fabricated and is undergoing tests. The test results of the 
optical transducer are also encouraging and an advanced system under design
is expected to  achieve similar sensitivity. The goal of the TIGA project is
to operate the detectors at  an overall noise of $E_{N} \cong 3 \hbar
\omega_{S}$. The sensitivity of the largest  sphere will then be $h_{min}
\cong 3 \times 10^{-22}$.

The GRAIL project is more ambitious. They plan to cool a 100 ton
copper-aluminum  sphere to 10 mK \cite{Frossati}. The sphere will be 3 m in
diameter and have the  fundamental frequency at 700 Hz. A four-mode
antenna-transducer system is envisioned  with a thin-film superconducting
inductive transducer \cite{Stevenson} as the last stage.

\begin{figure}[tbp]
\begin{center}
\leavevmode
\rotate[l]{\epsfxsize=3.5in\epsfbox{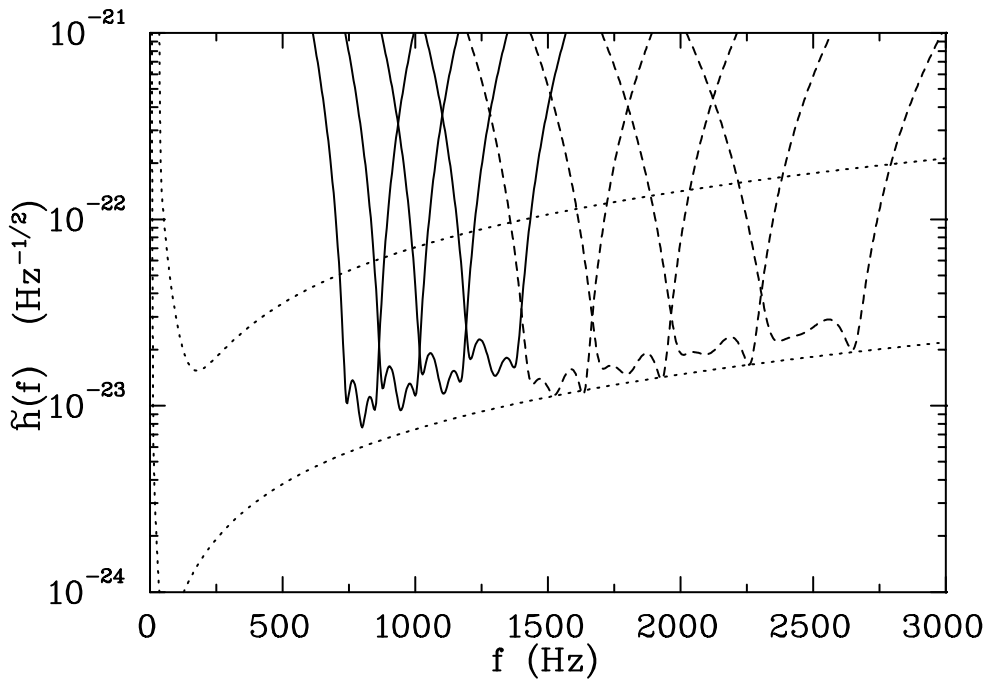}}
\end{center}
\caption{Strain spectrum of the four spherical antennas in the lowest and
second harmonic quadrupole modes in comparison with those of the first and
advanced LIGO}
\label{fig:Sensitivities}
\end{figure}

Figure \ref{fig:Sensitivities} shows the expected sensitivity of the TIGA
xylophone of four  spheres. The first four curves (solid lines) represent
the sensitivities of the lowest quadrupole modes  and the second four curves
(dashed lines) are those of the second harmonic quadrupole modes. The two 
dotted curves represent the expected sensitivities of the first and advanced
LIGO,  respectively. One can see that, although their bandwidths will be
somewhat restricted, the  spherical detectors will have 
sensitivities in their own bandwidths comparable to the long-baseline laser
interferometers. Therefore, the spherical antennas will complement the 
interferometers, at a modest increase in cost, with better resolution of the
source direction  and polarization as well as full-sky coverage.

\vskip 1cm \noindent {\bf VI. Gravitational-Wave Astronomy with Spheres} %
\vskip 0.7cm

What astrophysical sources will be observed by the fourth-generation
resonant-mass  detectors? It is difficult to tell because unexpected events
may be detected as has always happened  whenever new windows were opened for
optical, x-ray, infrared, and gamma-ray  astronomy. According to the present
theories, however, only a few GW sources have  been analyzed to the extent
that numerical waveforms have been computed. Although  these are not
completely relativistic calculations yet, they provide a useful guide for
the design  of the next-generation detectors.

Coalescing neutron star binaries are one type of source for which waveforms
have been  computed \cite{Centrella}. Harry {\it et al.} used these
waveforms to calculate the  expected signal-to-noise ratio for the TIGA
array \cite{Harry}. For two 1.4 $M_{\odot}$  neutron stars at the distance
of 15 Mpc (the approximate distance to the Virgo cluster), they found total
energy signal-to-noise ratios  (summed over the four spheres) of 25.8 and
0.51 for the inspiral and the coalescence  phase, respectively. This is to
be compared with the expected energy signal-to-noise  ratios of the first
LIGO of 38.0 and 0.010, respectively. For the inspiral phase, LIGO has 
greater sensitivity because of its wider bandwidth that extends down to 300
Hz. For the  coalescence phase, however, the TIGA xylophone has greater
sensitivity because the  signal is confined to frequencies above 1300 Hz.
Therefore, the spheres will have a better  chance to peer through to the
relativistic dynamics during the actual coalescence. The  expected event
rate for the above signals is less than 0.1 per year.

Another source for which waveforms have been computed is the bar mode
instability of a  rapidly rotating star formed during a supernova \cite{Smith}
Although the equatorial  radius $R_{E}$ of such a star is quite uncertain,
for $R_{E} = 20$ km, most of the signal  power is concentrated between 1 and
2 kHz where the spheres have the best sensitivity.  Thus, for a rotating
star of mass 1.4 $M_{\odot}$ and equatorial radius 20 km at the  distance of
1 Mpc, the TIGA array has the total energy signal-to-noise ratio of 51.8 in 
contrast to 0.88 for the first LIGO. The expected event rate for supernovae out to
that distance is about 0.1  per year.

Clearly, we need help from our theoretical colleagues to compute more
precise waveforms  and signal amplitudes for all plausible GW sources. For a
wide-band detector like a laser  interferometer, the exact waveform is even
more important since, without it, a template  cannot be constructed to
search for an event of a particular type.

Laser interferometers under construction and the spherical detectors on the
drawing  boards are expected to be sensitive enough to detect extragalactic
GW events. The next  decade may be one of the most exciting times in
gravitation and astrophysics as a new  window opens for astronomy. After so
much hard work pushing the frontiers of  materials, cryogenics, and sensor
technology for decades, with the sole purpose of  improving detectors for
nature's most elusive wave, it is about time to do some real  science with
these new devices!

\section*{Acknowledgements}

I would like to acknowledge useful discussions with Thomas Stevenson, Gregg
Harry,  Fred Wellstood, Insik Jin, and Charlie Misner. This work was
supported in part by the  National Science Foundation under grant
PHY93-12229.

\end{document}